\documentclass[,english,12pt]{iopart}

\usepackage{iopams}
\usepackage{amssymb}
\usepackage{setstack}
\usepackage{graphicx}
\usepackage{dcolumn}
\usepackage{bm}
\usepackage{dsfont}
\usepackage{amssymb}
\usepackage{subfigure}
\usepackage{t1enc,babel}

\newcommand{\bV     }{\mbox{\boldmath$V$}}
\newcommand{\bI     }{\mbox{\boldmath$I$}}

\newcommand{\bJ     }{\mbox{\boldmath$J$}}

\newcommand{\bu     }{\mbox{\boldmath$u$}}

\newcommand{\bA     }{\mbox{\boldmath$A$}}

\newcommand{\bS    }{\mbox{\boldmath$S$}}

\let\a=\alpha \let\b=\beta \let\g=\gamma \let\d=\delta
\let\e=\varepsilon   
\let\l=\lambda    \let\p=\pi
\let\s=\sigma \let\t=\tau

 \let\r=\rho

\newcommand{\beq}{\begin{equation}} \newcommand{\eeq}{\end{equation}}

\begin{document}

\title{Statistical mechanics of the spherical hierarchical model with random fields}

\author{Fernando L. Metz$^{1,2}$, Jacopo Rocchi$^{2}$, Pierfrancesco Urbani$^{3}$}
\address{$^1$Instituto de F\'{\i}sica, Universidade Federal do Rio Grande do Sul, Caixa Postal 15051, 91501-970 Porto Alegre, Brazil }
\address{$^2$  Dipartimento di Fisica, Sapienza Universit\'a di Roma, P.le A. Moro 2, I-00185 Roma, Italy}
\address{$^3$ IPhT, CEA/DSM-CNRS/URA 2306, CEA Saclay, F-91191 Gif-sur-Yvette Cedex, France }
\ead{fmetzfmetz@gmail.com}

\date{\today}  

\begin{abstract}
We study analytically the equilibrium properties of the spherical hierarchical model
in the presence of random fields. 
The expression for the critical line separating a paramagnetic
from a ferromagnetic phase is derived.
The critical exponents characterising this phase transition are computed analytically and compared 
with those of the corresponding $D$-dimensional short-range model, leading
to conclude that the usual mapping between one dimensional long-range models and $D$-dimensional short-range 
models holds exactly for this system, in contrast to models
with Ising spins. Moreover, the critical exponents of the 
pure model and those of the random field model satisfy a relationship that mimics the dimensional reduction 
rule. The absence of a spin-glass phase is strongly supported by the local stability
analysis of the replica symmetric saddle-point as well as by
an independent computation of the free-energy using a renormalization-like approach.
This latter result enlarges the class of random field models for which the spin-glass phase
has been recently ruled out.
\end{abstract}
\maketitle

\section{Introduction}

Random field models are ferromagnetic systems of spins with a random field at each site. 
They have been introduced by Imry and Ma \cite{imryma} and, despite many years of studies, their 
critical behaviour is still not completely understood, even in the simplest 
situation where the random fields are uncorrelated. 
The study of their critical behaviour is strictly related to a property 
called dimensional reduction, according to which the critical exponents of a random field 
model in $D$ dimensions is equivalent to those of the corresponding pure model in $D-2$ dimensions.
This property has been deeply analysed using different 
methods \cite{aha,you,parisi1979random}, and it has been
shown that dimensional reduction does not hold at low dimensions \cite{Imbrie}.
The reason for this behaviour is not entirely clear, and different scenarios
have been proposed to explain the critical behaviour of random field
models  (see \cite{tarjus2013critical} and references therein).

Some years ago it has been suggested that the main reason for the breakdown of
dimensional reduction 
could be the presence of an equilibrium spin-glass phase between a high temperature 
paramagnetic and a low temperature ferromagnetic phase \cite{de1995random}.
However, this intermediate spin-glass phase has 
been rigorously ruled out in a large class of random field 
models \cite{krzakala2010elusive,krzakala2011no}, which 
do present a breakdown of the dimensional
reduction property. These kind of models are defined
in terms of pairwise interacting spins placed on the vertices
of arbitrary graphs, where spins are of the Ising type \cite{krzakala2010elusive}. 
This result has been recently extended to the case of a scalar field theory \cite{krzakala2011no}.

Another interesting class of models consists of those in which
the couplings are arranged in an hierarchical structure.
The prototypical example is the
Dyson hierarchical model \cite{dyson1969existence}, introduced
a long time ago to study phase transitions in one-dimensional
models with interactions decaying as a power-law of the inter-site distance.
The Dyson hierarchical model has been intensively studied \cite{baker1972ising,baker1973spin, collet1978rg}, mostly because 
of its remarkable properties from the renormalization group point of view, which can be shown to be deeply connected with the 
approximate recursion formula derived by Wilson \cite{wilson1974renormalization,felder1987renormalization}, allowing
to efficiently compute critical exponents \cite{collet1978rg, kim1977critical}.

One of the main reasons why the Dyson hierarchical model
has attracted a lot of attention is that it can be used to investigate 
non-mean-field critical behaviour. In fact, short-range $D$-dimensional systems are well described by 
mean-field theory for $D$ large enough and a particular dimension separates mean-field from non-mean-field critical 
behaviour. Dyson hierarchical models exhibit qualitatively a similar phenomenology, with
the dimension $D$ replaced by an exponent $\tau$, the latter being responsible for controlling
the power-law decay of the interactions as a function of the inter-site distance.
Although this intuitive analogy has been recently exploited to study quenched disorder systems in 
the non-mean-field sector \cite{franz2009overlap, leuzzi2008dilute, katzgraber2009study, castellana2010hierarchical, 
castellana2010renormalization, castellana2011real,angelini2013ensemble, MonthusGarel1, banos2012correspondence, leuzzi2013imry, 
ParisiRoc, PhysRevE.89.032132, MariaChiara}, it is not clear to which extent there is 
a clear mapping between the spatial dimension $D$ and the exponent
$\tau$  \cite{banos2012correspondence,leuzzi2013imry, ParisiRoc, MariaChiara}.
Indeed, a strict mapping holds in the mean-field region, whereas it 
does not give satisfying results for low enough dimensions \cite{ParisiRoc, MariaChiara}.

Here we focus on a spherical version of the Dyson hierarchical model in the presence
of random fields, in which the spins are continuous variables and the phase space is constrained
to the surface of a sphere. The ferromagnetic spherical model was
introduced a long time ago by Berlin and Kac \cite{berlin1952spherical} and its hierarchical counterpart
has been considered in references \cite{mcguire1973spherical, ben2002phase}.
The main motivation
to study spherical models lies in their exactly solvable 
nature \cite{Baxter}, which renders a full analytical study of the critical properties
possible, even in the presence of quenched disorder. Although
the random field spherical model with short-range interactions
in $D$ dimensions has been previously studied \cite{hornreich1982thermodynamic}, the
Dyson hierarchical version constitutes an interesting testing ground 
to address at once different issues, such 
as the dimensional reduction problem, the existence of a spin-glass
phase and the aforementioned relationship between the critical properties of
short-range and long-range systems. We point out that the $D$-dimensional short-range
counterpart of our model always displays a non-zero Edwards-Anderson order-parameter \cite{hornreich1982thermodynamic}. This is probably 
not related to the
existence of a spin-glass phase, but simply reflects the presence of non-zero local
magnetisations due to the quenched random fields. 
A more refined analysis is certainly needed
in order to probe the existence of spin-glass states in the spherical model with random fields.

In this work we perform a thorough study of the equilibrium
properties of the spherical hierarchical model in the presence
of random fields, showing that the system undergoes a phase
transition between a paramagnetic and a ferromagnetic phase.
Exact analytical results for the critical
exponents are derived in the mean-field as well as in the non-mean-field regime.
By comparing our results with those of references \cite{mcguire1973spherical,hornreich1982thermodynamic}, we show that there is an exact
mapping between the critical exponents of the spherical hierarchical model and those of the
corresponding $D$-dimensional short-range system. 
Contrary to the case of Ising spins \cite{banos2012correspondence,leuzzi2013imry, ParisiRoc, MariaChiara}, such
mapping is exact here, even in the non-mean-field regime.
We also show that the critical exponents of the random field model 
and those of the corresponding pure model obey a certain relation, proposed 
in \cite {ParisiRoc}, which is analogous 
to the dimensional reduction property. In particular, this property holds for 
any value of $\tau$, in contrast to the hierarchical model with Ising spins \cite {ParisiRoc}, where
dimensional reduction breaks down in the non-mean-field regime.
The free-energy of the spherical hierarchical model is computed exactly using two different 
methods: a recursive approach \cite{collet1978rg}, based on the invariance of the
Hamiltonian under a renormalization-like transformation, and the standard replica
method \cite{Dominicis}.
Finally, we show that the replica symmetric saddle-point is locally
stable in the whole phase diagram, which strongly supports
the absence of a spin-glass phase. 

The rest of the paper is organised as follows.
In the next section we define the hierarchical spherical model
with random fields, while in section \ref{modelsolution}
we explain how the free-energy and the equation of state
for this model are derived using both the
recursive method and the replica method. The phase
diagram and the absence of a spin-glass phase are discussed
in section \ref{secphasediagram}. In section \ref{seccriticalexpo} we discuss the computation of the critical exponents and in section \ref{seccomments}
we comment on these results.
Some conclusions are presented in section \ref{secconclus}.

\section{The spherical hierarchical model with random fields} \label{themodelsec}

We study a one-dimensional model composed of $N= 2^n$ real-valued 
spins $\{ S_i \}_{i=1,\dots,N}$, with an external field 
$h_i = h + r_i$ acting on each spin $S_i$.
The quantity $h$ denotes the uniform part of the field, while $\{ r_i \}_{i=1,\dots,N}$ are drawn
independently from the Gaussian distribution 
\begin{equation}
p(r) = \frac{1}{\sqrt{2 \pi \sigma^2}} \exp{ \left( - \frac{r^2}{2 \sigma^2}  \right) } \,.
\label{Gauss}
\end{equation}
The system is governed by the following Hamiltonian
\begin{equation} 
\mathcal{H}(\bS) = - \frac{1}{2} \sum_{ij=1}^{N} J_{ij} S_i S_j - \sum_{i=1}^{N} h_i S_i \,,
\label{Hamilt1}
\end{equation}
where $J_{ij}$ denotes the coupling between $S_i$ and $S_j$.
The set of all couplings $\{ J_{ij} \}_{i,j=1,\dots,N}$  can be accommodated
in the symmetric interaction matrix $\bJ$, with 
$J_{ii}=0$ for $i=1,\dots,N$. 

Here we follow the original work of Dyson \cite{dyson1969existence} and
we consider a model where the couplings 
are organised in a hierarchical block structure. 
The hierarchy of interacting spins contains a total number
of $n$ levels, where $p=1$ and $p=n$ are, respectively, the lowest
and the highest level. The system
is divided into $2^{n-p}$ distinct groups or blocks of spins at a certain 
level $p$, each group containing $2^p$ mutually interacting spins. The coupling
between a pair of spins within a given block of level $p$ is 
defined as $b_p$. These definitions lead naturally to a matrix $\bJ$
where the off-diagonal elements are arranged in a
block structure, with blocks of dimension $1,2,2^2,\dots,2^{n-1}$. 
These matrix elements   
are given explicitly by $ J_{ij}=\sum_{p=1}^{k}  b_p \:$, where 
$k=1+\lfloor \log_2 |i-j| \rfloor$, and $\lfloor x \rfloor$ denotes the largest integer not greater than $x$.
\begin{figure}
\centering
\includegraphics[scale=0.7]{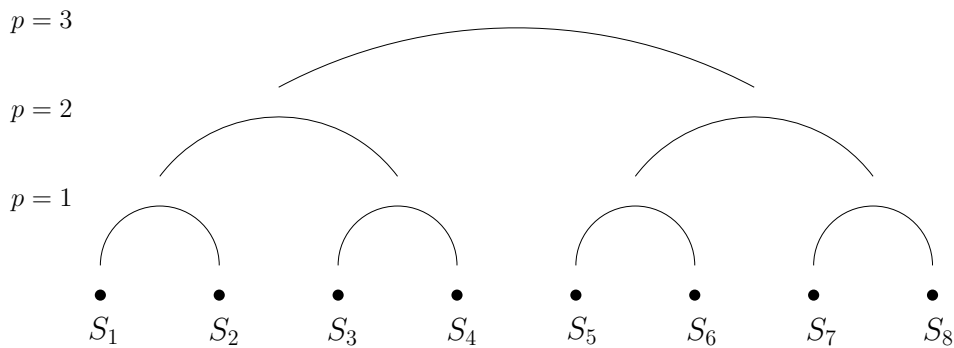}
\caption{Pictorial representation of the interactions between spins in 
the hierarchical model defined by the Hamiltonian of eq. (\ref{Hamilt})
with $n=3$ levels. The coupling between $S_i$ and $S_j$ is defined by $J_{ij}$. 
For instance, the interactions between $S_1$ and the other spins are explicitly given by $J_{12}=b_1+b_2+b_3$, 
$J_{13}= J_{14}= b_2+b_3$ and $J_{1,j}= b_3$, for $5 \leq j \leq 8$.
}
\label{fig:h_tree}
\end{figure}
Substituting the explicit form of $\bJ$ in eq. (\ref{Hamilt1}), the Hamiltonian reads
\begin{equation} 
\mathcal{H}(\bS) = - \frac{1}{2} \sum_{p=1}^{n} b_p \sum_{r=1}^{2^{n-p}} \left( \sum_{i=1}^{2^p} S_{(r-1)2^p + i}    \right)^2
+ \frac{A_n}{2} \sum_{i=1}^{2^n} S_i^2 - \sum_{i=1}^{2^n} h_i S_i \,,
\label{Hamilt}
\end{equation}
where $b_p =2^{-\tau p}$, with $\tau > 1$. As we will see below, the latter condition is required
to obtain a bounded spectrum for the matrix $\bJ$ in the limit $N \rightarrow \infty$.
The term including $A_n = \sum_{p=1}^{n} b_p$ removes the self-interactions of the model.
A schematic representation of the hierarchical block structure of the model
is displayed on figure \ref{fig:h_tree}.

The coupling
between two spins separated by a distance of $O(N)$ scales as $O(1/N^{\tau})$, and the hierarchical model exhibits the same
long-distance behaviour as a one-dimensional model with interactions
decaying as a power-law of the inter-site distance. 
By varying the exponent
$\tau$, this class of one-dimensional models interpolates between
systems with long-range or mean-field interactions and systems with 
short-range interactions, typical
of models defined on finite-dimensional lattices. 
The hierarchical arrangement of the couplings has an extra
advantage: the Hamiltonian preserves its structure under renormalization
transformations, which usually leads to exact iterative methods of
solution.

The matrix $\bJ$ has $n+1$ different eigenvalues
given by
\begin{equation}
\lambda_p^{(n)} = \frac{2^{-\tau n}  - 1 }{2^\tau - 1} + \frac{1 - 2^{- (\tau -1) (p-1)} }{2^{\tau - 1} - 1 }, \quad p = 1,\dots,n+1\,.
\label{eigen}
\end{equation}
For $p < n+1$, the eigenvalue $\lambda_p^{(n)}$ has a degeneracy factor of $2^{n-p}$, which 
comes from the symmetry between the blocks at level $p$. The largest 
eigenvalue, obtained by setting  $p=n+1$ in eq. (\ref{eigen}), has a degeneracy factor equal to $1$. 
In the thermodynamic limit, we can introduce the spectral density 
\beq
\r(\l)
=\sum_{p=1}^\infty2^{-p}\d(\l-\l_p)\:,
\label{DOS2}
\eeq
where $\l_p=\lim_{n\rightarrow \infty} \l_p^{(n)}$. The spectrum
of $\bJ$ is clearly bounded provided $\tau > 1$, with the largest eigenvalue
defined, for $n \rightarrow \infty$, as follows
\begin{equation}
\lambda_{\infty} = \lim_{p \rightarrow \infty} \lambda_p = \lambda_1 + \frac{1}{2^{\tau -1} - 1}\,,
\end{equation}
with $\lambda_{\infty}$ representing an accumulation point of
the spectrum, since the difference $\lambda_{p+1} - \lambda_{p}$ 
vanishes exponentially as a function of $p$. 

The partition function of the system in equilibrium at
temperature $T=\beta^{-1}$ reads
\begin{equation}
\mathcal{Z}_{N} = \int d \bS \,  \delta\left(N - |\bS|^2  \right) 
\exp{\left[ -\beta \mathcal{H}(\bS) \right] }
 \,,
\label{Z1}
\end{equation}
where $\bS = (S_1, \dots, S_N)$ and $d \bS = \prod_{i=1}^{N} d S_i$.
The Dirac delta function imposes the spherical 
constraint by restricting the above integral to the global configurations that fulfill $\sum_{i=1}^N S_i^2 = N$.  
Due to this spherical constraint, an
arbitrary real parameter $a$ can be introduced in $\mathcal{Z}_{N}$, which
allows us to rewrite eq. (\ref{Z1}) as follows
\begin{equation}
\mathcal{Z}_{N} = \int d \bS \int_{-\infty}^{\infty} \frac{ds}{2 \pi} \, 
\exp{\left[ -\beta \mathcal{H}(\bS) + \left(a + is \right) \left(N -  |\bS|^2  \right)    \right] }
 \,,
\label{part1}
\end{equation}
where we have used the Fourier integral representation of the
Dirac delta function. As will be clearer below, the introduction
of the regularizer $a$ is convenient to guarantee the convergence 
of the integral over $\bS$. 

Our primary aim consists in computing the free-energy
per spin in the thermodynamic limit
\begin{equation}
f = - \lim_{N \rightarrow \infty} \frac{1}{\beta N} \ln \mathcal{Z}_{N} \,,
\label{freeenerg}
\end{equation}
from which we have access to the thermodynamics of the model and, eventually, to its
critical behaviour.

\section{Solution of the model} \label{modelsolution}

In this section we present two different methods to compute
the partition function in the limit $N \rightarrow \infty$.
The recursive method explores the hierarchical structure
of the model and the trace over the spins is calculated
iteratively by making a sequence of renormalization-like
transformations on the partition function. 
This method is rather flexible, in the
sense that it allows to calculate the trace over the spins
for a single, finite instance of the random fields. In a subsequent
stage, one needs to employ the saddle-point method to  
evaluate $\ln \mathcal{Z}_{N}$ for $N \rightarrow \infty$, and the statistical properties
of the random fields become important.

The second approach is the well-known replica 
method \cite{Dominicis}, where the average of $\ln \mathcal{Z}_{N}$  over
the random fields is computed by replicating the system
$q$ times. In the replica setting, the limit  $q \rightarrow 0$
can be only performed by assuming a particular structure
for the order-parameters.
Here we show that the simplest assumption, namely the replica 
symmetric {\it ansatz}, yields the same averaged free-energy and 
the same equation of state as obtained through the recursive 
approach, provided the distribution of fields $p(r)$ is given by eq. (\ref{Gauss}).

\subsection{The recursive method}

In this section we show how to compute the trace over the spins 
using a simple change of integration variables combined with
the model hierarchical structure. Such approach 
has been employed to study hierarchical
models in the context of interacting spin systems (see \cite{JacopoTese} and references therein)
and Anderson localisation \cite{MonthusGarel,Metz2013, Metz2014}.

By substituting eq. (\ref{Hamilt}) in eq. (\ref{part1})
and choosing $a$ 
sufficiently large, we assure that the integral over $\bS$ in
eq. (\ref{part1}) is convergent, which allows us to interchange
the order of the $d \bS$ and $ds$ integrations 
\begin{equation}
\mathcal{Z}_{N} = \int_{-\infty}^{\infty} \frac{ds}{2 \pi} \, \exp{\left[  \left(a + is \right) N  \right]}
\, T_n (b_{1,\dots,n}, A_n,h_{1,\dots,2^n} \, | \, s) \,,
\label{Zdef1}
\end{equation}
where $T_n (b_{1,\dots,n}, A_n,h_{1,\dots,2^n} \, | \, s  )$ is the trace over the spins of a model with $n$ levels
\begin{eqnarray}
\fl
T_n (b_{1,\dots,n}, A_n,h_{1,\dots,2^n}\, | \, s ) &=& \int d \bS \, \exp{\left[  \beta  \sum_{i=1}^{2^n} h_i S_i   
- \left( a + \frac{\beta A_n}{2} + i s   \right) \sum_{i=1}^{2^n} S_i^2  \right]}, \nonumber \\
\fl
&\times&
\exp{\left[  \frac{\beta}{2} L_n (S_{1,\dots,2^n},b_{1,\dots,n})    \right]}
\label{Tdef}
\end{eqnarray}
and $L_n (S_{1,\dots,n},b_{1,\dots,n})$ encodes the hierarchical interactions of the Hamiltonian
\begin{equation}
L_n (S_{1,\dots,2^n},b_{1,\dots,n}) = \sum_{p=1}^{n} b_p \sum_{r=1}^{2^{n-p}} \left( \sum_{i=1}^{2^p} S_{(r-1)2^p + i}    \right)^2 \,.
\end{equation}
The shorthand notation $x_{1,\dots,\mathcal{S}} \equiv (x_1,\dots,x_{\mathcal{S}})$ has been
introduced to denote sets of variables.

The central idea consists in deriving a recursion equation
between the trace of a system with $n$ levels and the trace of a
system with $n-1$ levels, but with renormalized parameters.
This is achieved by making the following change of integration variables
in eq. (\ref{Tdef})
\begin{equation}
S_i^{\pm} = \frac{1}{\sqrt{2}} \left( S_{2i - 1} \pm  S_{2i}   \right), \quad i = 1,\dots,2^{n-1},
\label{changevar}
\end{equation}
which allows us to compute explicitly the Gaussian integrals 
over $\{ S_i^{-} \}_{i=1,\dots,2^{n-1}}$, provided
$a$ is chosen such that $a > - \beta A_n/2$. 
The fulfillment of the latter condition ensures the convergence 
of the integrals over $\{ S_i^{-} \}_{i=1,\dots,2^{n-1}}$. 
The number of degrees of freedom is reduced by one half
after integrating $\{ S_i^{-} \}_{i=1,\dots,2^{n-1}}$ out, 
and the variables $\{ S_i^{+} \}_{i=1,\dots,2^{n-1}}$ enter in the definition of 
a function $T_{n-1}$, which has the same formal structure 
as eq. (\ref{Tdef}), but with renormalized parameters.
Consequently, one can apply the above change of integration variables 
$\ell$ times in a consecutive way, obtaining the following
relation between $T_n(b_{1,\dots,n}, A_n,h_{1,\dots,2^n} \, | \, s)$ in the
original system and $T_{n-\ell} (b_{1,\dots,n-\ell}^{(\ell)}, A_n^{(\ell)},h_{1,\dots,2^{n-\ell}}^{(\ell)} \, | \, s)$
in a system with $n-\ell$ levels
\begin{eqnarray}
\fl
T_n (b_{1,\dots,n}, A_n,h_{1,\dots,2^n} \, | \, s) &=& T_{n-\ell} (b_{1,\dots,n-\ell}^{(\ell)}, A_n^{(\ell)},h_{1,\dots,2^{n-\ell}}^{(\ell)} \, | \, s) \nonumber \\
\fl
&\times&
\exp{\left[\frac{\beta^2}{8} \sum_{p=1}^{\ell} \frac{1}{\left(a + is  + \frac{\beta}{2}A_n^{(p-1)}  \right)}  \sum_{r=1}^{2^{n-p}}   
\left(  h_{2r - 1}^{(p-1)} -  h_{2r}^{(p-1)}    \right)^2   \right] } \nonumber \\
\fl
&\times&
\exp{\left[ - \frac{1}{2} \sum_{p=1}^{\ell} 2^{n-p} \ln{\left[ \frac{1}{\pi} \left( a + is + \frac{\beta}{2}A_n^{(p-1)}    \right) \right]}       \right]} \,, 
\label{eqT}
\end{eqnarray}
where
the renormalized parameters fulfill
\begin{eqnarray}
b_p^{(\ell)} &=& 2^{\ell} b_{p+\ell} \, \label{brenorm},  \\
A_n^{(\ell)} &=& A_n - \sum_{p=1}^{\ell} 2^p b_p \, , \label{eqA} \\
h_i^{(\ell)} &=& \frac{1}{2^{\frac{\ell}{2}}} \sum_{j=1}^{2^{\ell}} h_{2^{\ell} i + 1 - j} \, , \qquad i = 1,\dots,2^{n-\ell} \, \label{renormfield}.
\end{eqnarray}

By setting $\ell=n$ in eq. (\ref{eqT}), its right hand side depends
on the trace $T_0 (A_n^{(n)},h_{1}^{(n)} \, | \, s)$ of a renormalized single-spin problem.
Since  $T_0$ is defined in terms of a 
Gaussian integral over a single spin variable, this object can be computed in a straightforward way, leading
to an explicit expression for $T_n (b_{1,\dots,n}, A_n,h_{1,\dots,2^n} \, | \, s)$ as a function
of the renormalized parameters.
Substituting this explicit form of $T_n$ in eq. (\ref{Zdef1}), expressing the renormalized parameters
in terms of those of the original model through eqs. (\ref{brenorm}-\ref{renormfield}), and changing the integration
variable from $s$ to $z = (2/\beta) \left( a + i s - \beta \lambda_{n+1}^{(n)  } /2  \right)$,
we obtain
\begin{equation}
\mathcal{Z}_{N} = \frac{\beta}{4 \pi i} \int_{{\rm Re}z - i \infty}^{{\rm Re}z + i \infty} d z 
\exp{\left[ N \Phi_N \left(z \, | \, h_{1,\dots,2^n}  \right)   \right]} \,,
\label{Zint}
\end{equation}
with
\begin{eqnarray}
\fl
& \Phi_N& \left(z \, | \, h_{1,\dots,2^n}  \right) = \frac{1}{2} \ln{\left( \frac{2 \pi}{\beta}  \right)} + \frac{\beta}{2} \left( z + \lambda_{n+1}^{(n)}  \right)
- \frac{1}{2}  \sum_{p=1}^{n} \frac{1}{2^{p}} \ln{\left( z + \lambda_{n+1}^{(n)} - \lambda_{p}^{(n)}     \right)}   \nonumber \\
\fl
&-&  \frac{1}{2N} \ln{z} + \frac{\beta}{2 z} \left( \frac{1}{N} \sum_{i=1}^{N} h_i  \right)^2  \nonumber \\
\fl
&+& \frac{\beta}{2 N} \sum_{p=1}^{n} \frac{1 }{2^{p} \left( z + \lambda_{n+1}^{(n)} - \lambda_{p}^{(n)}     \right)   }
\sum_{r=1}^{2^{n-p}} \left[ \sum_{i=1}^{2^{p-1}} \left(  h_{(2 r -1) 2^{p-1} + 1 -i } - h_{r 2^{p} + 1 -i }      \right)    \right]^2\, .
\label{phifunc}
\end{eqnarray}
All Gaussian integrals over the spin variables, involved in the derivation of 
eqs. (\ref{Zint}) and (\ref{phifunc}),
are convergent provided $a > -\beta A_n^{(n)}/2$. From eqs. (\ref{eigen}) and
(\ref{eqA}), one can show that $A_n^{(\ell)} = - \lambda_{\ell+1}^{(n)}$ ($\ell=0,\dots,n$). 
The condition $a > \beta \lambda_{n+1}^{(n)}/2$ is also found 
in the approach based on the diagonalization
of $\bJ$ \cite{Baxter}.
We point out that eqs. (\ref{Zint}) and (\ref{phifunc}) are completely
general in the sense that they hold for a finite realisation
of the random parts $r_{1},\dots,r_{2^n}$ of the local fields $h_{1},\dots,h_{2^n}$, independently
of their distribution $p(r)$.

In order to make further progress, let us assume that $\{ r_i \}_{i=1,\dots,N}$ are independently drawn
from eq. (\ref{Gauss}).
In this case, we have checked numerically that the standard deviation of the
random part of eq. (\ref{phifunc}) is of $O(1/\sqrt{N})$ for
$N \gg 1$. Hence, the function $\Phi_N \left(z \, | \, h_{1,\dots,2^n}  \right)$
is a self-averaging quantity, \emph{i.e.}, it converges, in the limit $N \rightarrow \infty$, to its average
value 
\begin{eqnarray}
\Phi \left(z \right) &=& \frac{1}{2} \ln{\left( \frac{2 \pi}{\beta}  \right)} + \frac{\beta}{2} \left( z + \lambda_{\infty}  \right)
+ \frac{\beta h^2}{2 z}  - g(z) + \beta \sigma^2 \frac{\partial g(z)}{\partial z} \,,
\label{phifunc1}
\end{eqnarray}
where $g(z)$ is written in terms of the
density of eigenvalues given in eq. (\ref{DOS2}):
\begin{equation}
g(z) = \frac{1}{2} \int d \lambda \, \rho(\lambda) \ln{\left( z + \lambda_{\infty} - \lambda  \right)}\,.
\label{gz}
\end{equation}
It follows that the integral of eq. (\ref{Zint}) can
be solved, in the limit $N \rightarrow \infty$, through the saddle-point method, leading to the free-energy
\begin{equation}
f (z) = - \frac{1}{\beta} \Phi \left(z \right)\,,
\label{free}
\end{equation}
with the order-parameter $z$ satisfying the saddle-point equation
\begin{equation}
\frac{\beta}{2} \left( 1 -  \frac{ h^2}{z^2} \right) = \frac{\partial g(z)}{\partial z} - \beta \sigma^2 \frac{\partial^2 g(z)}{\partial^2 z}\,.
\label{eqstate}
\end{equation}
The magnetisation per spin $m$ is obtained from
the derivative of eq. (\ref{free}) with respect to $h$ 
\begin{equation}
m = \frac{h}{z}.
\label{eqmzh}
\end{equation}
The substitution of $z = h/m$ in eq. (\ref{eqstate}) yields
the equation of state for this  model. In the next subsection we
explain how the same results are derived  using the replica method.

\subsection{The replica method}

In the replica method, the average of the free-energy over
the quenched disorder
is computed using the following identity
\begin{equation}
\overline{ \ln \mathcal{Z}_{N} }  = \lim_{q \rightarrow 0} \frac{\partial}{\partial q} \overline{ \left( \mathcal{Z}_{N} \right)^{q} } \,,
\label{repl1}
\end{equation}
where $\overline{( \dots ) }$ denotes the average over the distribution
of the random fields. 
The strategy consists in calculating $\overline{ \left( \mathcal{Z}_{N} \right)^{q} }$ 
for integer $q$, which corresponds to averaging the product of the partition functions
of $q$ identical copies or replicas of the system. After the thermodynamic
limit is performed, $q$ is continued analytically to real values and finally to zero. 
In the problem at hand, we will show that the  
replica symmetric (RS) {\it ansatz} for the saddle-point leads, in the limit $q \rightarrow 0$, to the same free-energy as that 
computed with the recursive method.

Substituting eq. (\ref{Hamilt1}) in eq. (\ref{part1}) and performing the
average of the replicated partition function $\left( \mathcal{Z}_{N} \right)^{q}$
over the random variables $\{ r_i \}_{i=1,\dots,N}$, we obtain
\begin{eqnarray}
\fl
\overline{ \left( \mathcal{Z}_{N} \right)^{q} } &=& \int \left( \prod_{\alpha=1}^{q} d \bS^{\alpha}   \right)
\int  \left( \prod_{\alpha=1}^{q} \frac{d s_{\alpha}}{2 \pi}  \right)
\exp{\left[ N \sum_{\alpha=1}^{q} \left( a + i s_{\alpha}  \right)   
-\sum_{\alpha=1}^{q} (\bS^{\alpha})^T . \bV(s_{\alpha}) \bS^{\alpha}  \right]
} \nonumber \\
\fl 
&\times& \exp{\left[ 
\beta \sum_{\alpha=1}^{q} \bu^T .  \bS^{\alpha}
+ \frac{\beta^2 \sigma^2}{2} 
\sum_{\alpha, \beta=1}^{q}  (\bS^{\alpha})^T . \bS^{\beta} \right] } ,
\label{eqd}
\end{eqnarray}
where $(\bS^{\alpha})^T = (S_{1}^{\alpha},\dots,S_{N}^{\alpha})$ is the global
state vector in a given replica $\alpha$, while $\bu^T = (h,\dots,h)$ is the
$N$-dimensional vector including the uniform part of the external fields.
The matrix $\bV(s_{\alpha})$ is defined as follows
\begin{equation}
\bV(s_{\alpha}) = \left( a + i s_{\alpha} \right) \bI - \frac{\beta}{2} \bJ \,,
\end{equation}
with $\bI$ denoting the $N \times N$ identity matrix. 

Let us define 
the set of normalised eigenvectors and eigenvalues of $\bV(s_{\alpha})$
by $\{ \bphi_{k} \}_{k=1,\dots,N}$ and $\{ v_k(s_{\alpha}) \}_{k=1,\dots,N} $, respectively. 
The eigenvectors are independent of the replica index $\alpha$, since
they are the same as the eigenvectors of $\bJ$.
The insertion of the completeness relation for $\{ \bphi_{k} \}_{k=1,\dots,N}$ in
each term of eq. (\ref{eqd}) yields the following expression

\begin{eqnarray}
\fl
\overline{ \left( \mathcal{Z}_{N} \right)^{q} } &=& \int \left( \prod_{\alpha=1}^{q} \prod_{k=1}^{N} d P_{k}^{\alpha}   \right)
\int  \left( \prod_{\alpha=1}^{q} \frac{d s_{\alpha}}{2 \pi} \right)   \exp{\left[ N \sum_{\alpha=1}^q   \left( a + i s_{\alpha}  \right) \right] }   
  \nonumber \\
\fl 
&\times& \prod_{k=1}^{N} \exp{\left[ -\frac{1}{2} \sum_{\alpha, \beta=1}^{q}   P_{k}^{\alpha}  A_k^{\alpha \beta}(s_{\alpha})   P_{k}^{\beta} \,
+ \beta \, y_k  \sum_{\alpha=1}^{q}  P_{k}^{\alpha}    \right] }\,,
\label{eqZ1}
\end{eqnarray}
where the scalar projections $P_{k}^{\alpha} =  (\bS^{\alpha})^T . \bphi_k$ and $y_k = \bu^T . \bphi_k$
onto the eigenvectors $\{ \bphi_k \}_{k=1,\dots,N}$ have been introduced. The elements
of the $q$-dimensional matrix $\bA_k(s_{1,\dots,q})$ in the replica space
are given by
\begin{eqnarray}
A_k^{\alpha \beta}(s_{\alpha}) = 2 v_k(s_{\alpha}) \delta_{\alpha \beta} - \beta^2 \sigma^2 \,.
\label{matAk}
\end{eqnarray}
As usual in the replica method, the original model with quenched random fields
has been converted 
in a pure system composed of replicated variables $P^{1}_k,\dots,P^{q}_k$ that interact through the off-diagonal 
elements of $\bA_k(s_{1,\dots,q})$, as can be noted
from eq. (\ref{eqZ1}).
By assuming $a > \beta \lambda_{n+1}^{(n)} /2 + \beta^2 \sigma^2 / 2 $, we can
integrate over $P_{k}^{\alpha}$ and derive
\begin{equation}
\overline{ \left( \mathcal{Z}_{N} \right)^{q} } = \left( 2 \pi \right)^{\frac{N q}{2}} 
\int  \left( \prod_{\alpha=1}^{q} \frac{d s_{\alpha}}{2 \pi} \right) \exp{\left[ N  W\left(s_{1,\dots,q}\right) \right] }\,,  
\label{eqZ1a}
\end{equation}
where we have introduced the  action in the replica space
\begin{equation}
\fl
W(s_{1,\dots,q})=\sum_{\a=1}^q(a+is_{\alpha})  - \frac{1}{2N} \sum_{k=1}^N\ln \det \bA_k (s_{1,\dots,q})  
+ \frac{\beta^2}{2N}\sum_{k=1}^N y_k^2\sum_{\a,\b=1}^q  \left[ \bA^{-1}_k(s_{1,\dots,q}) \right]_{\alpha \beta} .
\label{actionrepl}
\end{equation}

In order to obtain the free-energy  we
make the replica symmetric {\it ansatz}, i.e.,
we assume that $s_{\alpha} = s$ for $\alpha=1,\dots,q$, from which 
one can show that
\begin{eqnarray}
&&\det{  \bA_k(s) } = \left[ 2 v_k(s)  \right]^q \left[ 1 - \frac{q \beta^2 \sigma^2}{2 v_k(s) }    \right]\,, \nonumber \\
&&\left[\bA^{-1}_k(s)\right]_{\a\b}=\frac{1}{2v_k(s)}\delta_{\a\b}+\frac{\b^2\s^2}{2v_k(s) \left[2v_k(s)-q\b^2\s^2 \right]}\,. \label{inverse_matrix}
\end{eqnarray}
The correctness of the RS assumption will be justified through a local stability analysis in the following section.
The insertion of the above two equations in eq. (\ref{eqZ1a}) reads
\begin{eqnarray}
\fl
\overline{ \left( \mathcal{Z}_{N} \right)^{q} } &=& \left( 2 \pi \right)^{\frac{N q}{2}} 
\int  \frac{d s}{2 \pi}  
\exp{\left[ N q \left( a + i s  \right)  
+  \frac{ \beta^2 q}{2} \sum_{k=1}^N \frac{y_k^2}{2v_k(s)-q \beta^2 \sigma^2 }    - \frac{q}{2} \sum_{k=1}^{N} \ln{\left[ 2 v_k (s)  \right]}
\right]} \nonumber \\
\fl 
&\times& \exp{\left[ - \frac{1}{2}  \sum_{k=1}^{N} \ln{\left(   1 - \frac{q \beta^2 \sigma^2  }{ 2 v_k (s)  }     \right)}    \right] }\,.
\label{eqZ1ab}
\end{eqnarray}
By noting that $\bu = \sqrt{N} h \bphi_N$, 
with $\bphi_N= N^{-1/2}(1,\dots,1)$ representing the uniform
eigenvector of $\bJ$ corresponding to the largest eigenvalue $\lambda_{n+1}^{(n)}$, we 
have that $y_k =  \sqrt{N} h \delta_{N,k}$, which
follows from the orthogonality among the eigenvectors.
This implies that only the term with $k=N$ survives in the
contribution involving $\{ y_{k}^{2} \}_{k=1,\dots,N}$.
The last step consists in inserting the explicit
form of $\{ v_k (s) \}_{k=1,\dots,N}$, given in terms of the eigenvalues
$\{ \lambda_{p}^{(n)} \}_{p=1,\dots,n+1}$ of $\bJ$, with the corresponding
degeneracy factors as defined just below eq. (\ref{eigen}), and to make 
the change of integration variable $z = (2 / \beta) \left( a + i s - \beta \lambda_{n+1}^{(n)} /2  \right)$, to obtain
\begin{equation}
\overline{ \left( \mathcal{Z}_{N} \right)^{q} } = \frac{\beta}{4 \pi i} \int_{{\rm Re}z - i \infty}^{{\rm Re}z + i \infty} d z 
\exp{\left[ N \Phi^N_{q} \left(z \right)   \right]} \,.
\label{Zintrepl}
\end{equation}
In the  limit $N \rightarrow \infty$, the sums $\sum_{k=1}^{N} (\dots)$ in eq. (\ref{eqZ1ab}) may be replaced by 
averages with the density of eigenvalues $\rho(\lambda)$, and the function $\Phi^N_{q} \left(z \right)$ converges
to the following well-defined limit 
\begin{equation}
\Phi_{q} \left(z \right) = \frac{q}{2} \ln{\left( \frac{2 \pi}{\beta}   \right) } + \frac{q \beta}{2} \left( z + \lambda_{\infty} \right) +
\frac{q \beta h^2  }{2 \left( z - q \beta \sigma^2    \right)} - q g(z) - g_q(z)\,,
\label{phiq}
\end{equation}
where $g(z)$ is defined by eq. (\ref{gz}), and $g_q(z)$ reads
\begin{equation}
g_q(z) = \frac{1}{2} \int d \lambda \rho(\lambda) 
\ln{\left[1 - \frac{q \beta \sigma^2 }{\left( z + \lambda_{\infty} - \lambda   \right) }   \right]}\,.
\end{equation}
The integral in eq. (\ref{Zintrepl}) is computed, in the limit $N \rightarrow \infty$, through the saddle-point
method, and the free-energy reads
\begin{equation}
f (z) =-\frac{1}{N \beta} \overline{ \ln{\mathcal{Z}_{N}} } = - \frac{1}{\beta} \lim_{q \rightarrow 0} 
\frac{\partial \Phi_{q} \left(z \right)}{\partial q} = -\frac{1}{\beta} \Phi \left(z \right)\:,
\label{freenerepit}
\end{equation}
where $\Phi \left(z \right)$ is defined by eq. (\ref{phifunc1}) and $z$ fulfills
the  saddle-point equation (\ref{eqstate}). 
The magnetization is computed in the replica method as follows
\begin{equation}
m =\frac{1}{N \beta}\frac{\partial}{\partial h} \overline{ {\ln \mathcal{Z}_N} } 
=\frac{1}{\beta }\frac{\partial }{\partial h} \lim_{q\rightarrow0}\frac{\partial\Phi_{q} \left(z \right)}{\partial q} = \frac{h}{z} .
\label{magreplica}
\end{equation}
Equations (\ref{freenerepit}) and (\ref{magreplica}) are the same as those
derived in the previous section by means of the recursive approach.


\section{Phase diagram and the stability of the RS saddle-point} \label{secphasediagram}

In this section we discuss the properties of the equation of state and the local stability of the RS 
assumption of the previous section. These studies allow us to provide
a rather complete characterisation of the phase diagram.


\subsection{Phase diagram}
The thermodynamics of the model is governed 
by two parameters: $\beta$ and $\sigma$. In order to study
the phase diagram,
we express the free energy, given in eq. (\ref{freenerepit}), in terms of the magnetisation $m=h/z$  
\begin{equation}
f(m,h)=f_{pure}(m,h)-\sigma^2 g' \left(\frac{h}{m}\right),
\label{freeenerffff}
\end{equation}
where $g'(z) = \frac{\partial g(z)}{\partial z}$, and  $f_{pure}(m,h)$ is the free-energy $f(m,h)$ calculated at $\sigma=0$:
\begin{equation}
f_{pure}(m,h)=-\frac{1}{2 \beta} \ln \left(\frac{2\pi}{\beta} \right) -\frac{1}{2} \left( \frac{h}{m}+\lambda_{\infty} \right)-\frac{hm}{2}
+\frac{1}{\beta}g\left(\frac{h}{m}\right)\:.
\label{freefinal}
\end{equation}

The equation of state is obtained by writing the saddle-point equation (\ref{eqstate}) in terms
of $m$ and $h$
\begin{equation}
1-m^2=\frac{2}{\beta}g'\left(\frac{h}{m}\right) - 2\sigma^2 g''\left(\frac{h}{m}\right)\:. 
\label{eqstatfinalm}
\end{equation}
In the ferromagnetic phase, where $|m| > 0$, we
can safely send $h$ to zero in eq. (\ref{eqstatfinalm}), obtaining
an equation in terms of $g'( 0)$ and  $g''( 0)$, which
depends on the magnetisation $m$. 
The resulting equation is satisfied at the critical
values $\beta_c$ and $\sigma_c$, by definition, if $m=0$. Thus, the critical line separating 
the ferromagnetic from the paramagnetic phase on the $\sigma-\beta$ reads
\begin{equation}
\frac{\beta_c}{2}=g'(0)-\sigma_c^2 \beta_c g''(0)\:.
\label{eq:jac_eqforthecrtil}
\end{equation}

Equation (\ref{eq:jac_eqforthecrtil}) is a self-consistent equation that gives a value of the critical variance $\sigma^2_c$ for each value of the 
temperature $T_c=1/\beta_c$, and vice versa.
However, the existence of a solution of this equation depends on the value of $\t$, which
governs the analyticity properties of the functions
$g'(z)$ and $g''(z)$ around $z=0$. In the appendix, we show
that $g'(0)$ and $g''(0)$ are finite provided $\t<2$
and $\t<3/2$, respectively.
This means that, for $\t<3/2$, a solution $\s_c^2(T)\geq0$ exists, and there 
is a critical line that connects the zero temperature critical point $\s_c^2(0)$ with the finite 
temperature critical point $\b_c(\s^2=0)=2g'(0)$ of the pure model. For $\s^2>0$, the critical 
line shrinks to the $T$ axis when $\t\to 3/2$, so 
that no phase transition is present for $\t\geq3/2$. For $\s^2=0$, the model has a finite 
critical temperature which goes to zero when $\t\to 2$. 

\begin{figure}
\centering
\includegraphics[scale=1.4]{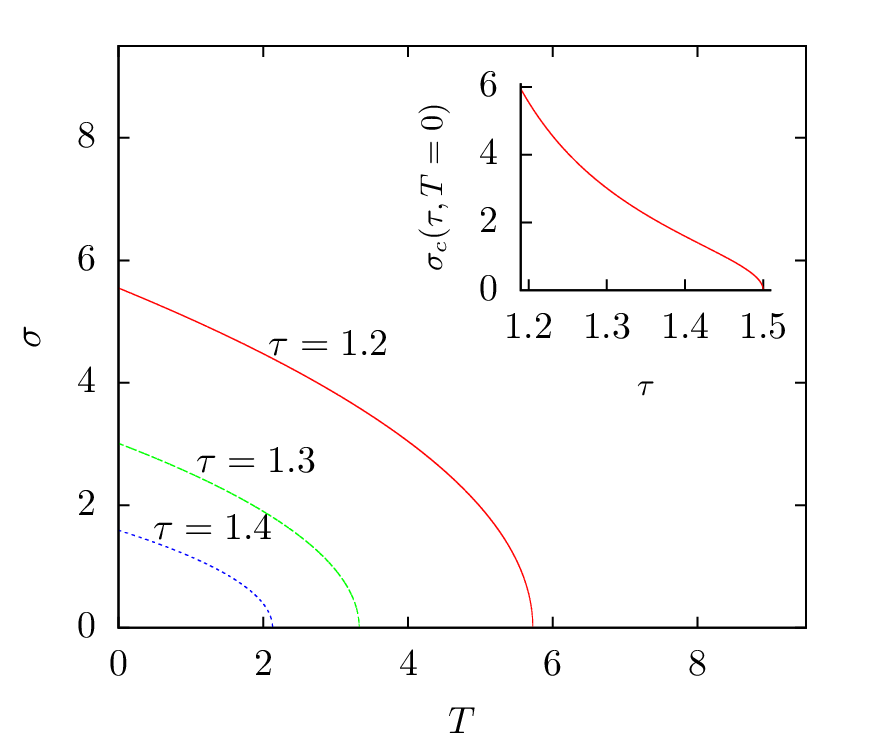}
\caption{Phase diagram of the hierarchical spherical model with random fields 
for different values of $\tau$.
The system is in a paramagnetic or in a ferromagnetic phase depending if the parameters 
are chosen above or below
the critical line, respectively. 
The inset displays the critical standard deviation of the random fields as function of $\tau$
at zero temperature.}
\label{fig:P_D}
\end{figure}

The phase diagram is reported in Fig. \ref{fig:P_D}.
Below the critical line, the system is in a ferromagnetic phase, where the magnetisation is 
different from zero. Although we have $m=0$ above the critical line, an intermediate
spin-glass phase can not be excluded.
It has been shown rigorously in \cite{krzakala2010elusive,krzakala2011no} that certain random field models 
undergo a transition between a paramagnetic and a ferromagnetic phase, without the appearance of any intermediate spin-glass phase.
Despite these results, we cannot exclude {\it a priori} an intermediate spin-glass phase, since
the present model does not belong to the class of systems studied in \cite{krzakala2010elusive,krzakala2011no}.
In the following we will show that the replica symmetric solution is locally stable.
This result, combined with the fact that we obtain the same free-energy
through the recursive and the replica method, strongly supports
the absence of a spin-glass phase.       


\subsection{Stability of the RS {\it ansatz}}
In this section we study the stability properties of the matrix that governs the small fluctuations around the RS saddle-point.
This is important in order to see if there is the possibility of a spin-glass phase with 
replica symmetry breaking.
Let us consider the action defined by eq. (\ref{actionrepl}).
The small fluctuations around the RS saddle-point are controlled
by the eigenvalues of the stability matrix \cite{Almeida}
\beq
M_{\a\b}=\frac{\partial^2 W}{\partial s_\a\partial s_\b} \Big{|}_{RS} \,,
\eeq
where  $(\dots )|_{RS}$ means that the second derivatives are calculated 
at the RS saddle-point, characterised by $s_{\alpha}=s$ for $\alpha=1,\dots,q$.

Equation (\ref{matAk}), combined with the explicit form for the 
eigenvalues $v_{k}(s_\alpha)$, allows us to show that
$\partial \left(\bA_k\right)_{\a\b}/\partial s_\g=2 i \delta_{\a\b}\delta_{\a\g}$.
By using this result and the following general 
relations
\beq
\fl
\frac{\partial \left(\bA_k^{-1}\right)_{\a\b}}{\partial (\bA_k)_{\g \delta}}=-\left(\bA_k^{-1}\right)_{\a \g}\left(\bA_k^{-1}\right)_{\delta \b},
\qquad
\frac{\partial}{\partial {(\bA_k)_{\a\b}}} \ln \det {\bA_k}=\left(\bA_k^{-1}\right)_{\b\a}\:,
\nonumber 
\eeq
one can derive the expression for the second derivative
\begin{eqnarray}
\fl
 \frac{\partial^2 W}{\partial s_\g\partial s_\omega} &=& -\frac{2}{N } \sum_{k=1}^{N} 
\left({\bA_k^{-1}}\right)_{\g\omega} \left({\bA_k^{-1}}\right)_{\omega\g} - \frac{2 \beta^2}{N} \sum_{k=1}^N y_k^2 
\sum_{\alpha,\beta=1}^{q}  \left({\bA_k^{-1}}\right)_{\a\omega}  \left({\bA_k^{-1}}\right)_{\omega\g}  \left({\bA_k^{-1}}\right)_{\g\b}  \nonumber \\
\fl
&-& \frac{2}{N } \sum_{k=1}^{N} 
\sum_{\alpha,\beta=1}^{q} \left({\bA_k^{-1}}\right)_{\a\g}  \left({\bA_k^{-1}}\right)_{\g\omega}  \left({\bA_k^{-1}}\right)_{\omega\b} \,.
\end{eqnarray}
We need to compute the stability matrix at the RS solution. In this case, 
the inverse $\bA_k^{-1}$ is given by eq. (\ref{inverse_matrix}), and
the elements of the stability matrix at the RS saddle-point
assume the form $M_{\a\b}=M_1\d_{\a\b}+M_2$, where 
\beq
M_1=-\frac{1}{2N}\sum_{k=1}^NA^{(k)}_D\left[2\beta^2 y_k^2\left(A^{(k)}_D+qA^{(k)}_F\right)^2+A^{(k)}_D+2A^{(k)}_F\right],
\label{m111111}
\eeq
\beq
M_2=-\frac{1}{2N}\sum_{k=1}^NA^{(k)}_F\left[2\beta^2 y_k^2\left(A^{(k)}_D+qA^{(k)}_F\right)^2+\left(A^{(k)}_F\right)^2\right],
\label{m222222}
\eeq
with the coefficients
\beq
A^{(k)}_D=\frac{1}{2v_k(s)} , \qquad
A^{(k)}_F=\frac{\b^2\s^2}{2v_k(s)\left(2v_k(s)-q\b^2\s^2\right)} ,
\label{akddddd}
\eeq
which are the diagonal and the off-diagonal parts of $\left({\bA}^{-1}\right)_{\a\b}$, as can be 
seen from eq. (\ref{inverse_matrix}).

The eigenvalues of the stability matrix are given by
\beq
\l_1=M_1+qM_2 \,, \qquad
\l_2=M_1\,.
\eeq
The degeneracy of $\l_1$ is $1$ while the degeneracy of $\l_2$ is $q-1$. 
We want to compute, in the limit $q\to 0$, the eigenvalues 
$\lambda_1$ and $\lambda_2$ at the 
solution $z$ of the saddle-point equation (\ref{eqstate}). 
In the limit $h \rightarrow 0$ of vanishing external field, $z$ tends
to a value $z_0$ and one can show that $z_0 = 0$ in the ferromagnetic phase, while
$z_0 > 0$ in the paramagnetic phase (see the
next section).
Thus, in the absence of external field, the eigenvalues $v_k(s)$, expressed
as a function of $z_0$, are given by
$v_k(z_0)=(\b/2)(z_0+ \l_{n+1}^{(n)}  -\l_{k}^{(n)})\geq 0$, which implies that
$A^{(k)}_D$ and $A^{(k)}_F$ are positive quantities for  $q \rightarrow 0$.
As a consequence, the functions $M_1$ and $M_2$, defined by eqs. (\ref{m111111}) and (\ref{m222222}), are negative
in the limit $q \rightarrow 0$. Thus, the 
eigenvalues $\l_1$ and $\l_2$ of the stability matrix are negative in the whole
phase diagram, implying the local stability of the RS solution.


\section{Critical exponents} \label{seccriticalexpo}

In order to compute the critical exponents, we need to study eq. (\ref{eqstate})
close to the critical line.
Let us define $K=\beta/2$.
As $h\rightarrow 0^+$, the l.h.s. of eq. (\ref{eqstate}),
$L_h(z)=K\left(1-h^2/z^2\right)$, is more 
and more close to the constant value $K$, while the r.h.s. $R_h(z)=g'(z)- 2 K \sigma^2 g''(z)$ is a smoothly decaying function. 
Let us define $z_0=\lim_{h\rightarrow 0} z^{*}(h)$, where $z^{*}(h)$ is the intersection
point between $L_h(z)$ and $R_h(z)$.
The function $R_{h\to 0^+}(z=0)$ determines whether or not $m>0$: for $R_{h\to 0^+}(z=0)>K$, $z_0$ is finite and positive, and thus $m=0$, while 
for $R_{h\to 0^+}(z=0)<K$, $z_0$ is zero, and thus $m>0$. 
We are going to use eq. (\ref{eqstate}) to compute the critical exponents. 

In this model the critical 
point may be approached following different directions in the $(T,\s)$ phase diagram.
We limit ourselves to discuss the situations when we vary $\sigma$ at a fixed value of $T$, and vice versa.
The critical behaviour does not depend on this choice.
Since the critical exponents depend crucially on the behaviour of the solution $z_0$ of eq. (\ref{eqstate})
close to the critical line, we first study such behavior coming from the 
paramagnetic phase and approaching the critical line. 

Let us first study the case when $T$ is fixed and we vary $\sigma^2$. The critical value $\sigma_c^{2}(T)$ reads
\begin{equation}
\sigma^2_c(T)
=\frac{g'(0)-K}{2 K g''(0)}\:.
\label{eq:jac_sigmact1}
\end{equation} 
In the paramagnetic phase, we have that $z_0 > 0$ and the limit $h\to 0$
can be safely taken in eq. (\ref{eqstate}) to obtain
\beq
K=g'(z_0)-2 K \s^2 g''(z_0)\:.
\eeq
Close to the critical line, we have to study the behaviour of $g'(z_0)$ and $g''(z_0)$ for small $z_0$. 
We focus on the region where $\t<3/2$, which is the relevant one for the random field model, namely for $\s^2>0$.
The behaviour of $z_0$ depends crucially on the value of $\t$. In fact, by using
the results presented in the appendix, we can show that for $\t <4/3$ we have
\beq
z_0\sim(\s^2-\s_c^2(T)) \,,
\eeq
while for $\t>4/3$ the equation of state becomes
\beq
0=(\s^2-\s_c^2(T))g''(0)+\s^2 C_1 z_0^{\frac{3-2\t}{\t-1}}\,,
\eeq
with $C_1$ denoting an unimportant constant.
For $\t>4/3$, $g'''(0)$ does not exist, but $g''(z)$ is given by $g''(0)+ C_1 z^{\frac{3-2\t}{\t-1}}$ 
(at the leading order in $z\ll1$), because its derivative $g'''(z)$ is proportional to  $z^{\frac{4-3\t}{\t-1}}$.
For $\s\to \s_c(T)$, this means that
\begin{equation}
z_0 \sim \left\{ \begin{array}{rl}
 &\left[\s-\s_c(T) \right] \mbox{ $\qquad$ if $ \tau<\frac{4}{3}$ \,,} \\
 &\left[\s-\s_c(T)\right]^{\frac{\tau-1}{3-2\tau}} \mbox{ $\:\:$ if $ \frac{4}{3}<\tau<\frac{3}{2}$\,.}
       \end{array}  \right.
\label{eq:primaconfrr}
\end{equation}
By repeating exactly the same analysis for fixed $\sigma^2$, one can show that, for
$K\to K_c(\s)$, $z_0$ behaves as follows
\begin{equation}
z_0 \sim \left\{ \begin{array}{rl}
 &\left[K-K_c(\sigma) \right] \mbox{ $\qquad$ if $ \tau<\frac{4}{3}$\,,} \\
 &\left[K-K_c(\sigma)\right]^{\frac{\tau-1}{3-2\tau}} \mbox{ $\:\:$ if $ \frac{4}{3}<\tau<\frac{3}{2}$\,.}
       \end{array}  \right. 
              \label{eq:secondaconfrr}
\end{equation}
For the pure model, where $\sigma^2=0$, the r.h.s. of eq. (\ref{eqstate}) is well 
defined in a larger domain $\tau<2$ and, in a similar way, we obtain, for $K\to K_c(\s=0)$, the following
result
\begin{equation}
z_0 \sim \left\{ \begin{array}{rl}
 &\left[K-K_c(\sigma=0) \right] \mbox{ $\qquad$ if $ \tau<\frac{3}{2}$\,,} \\
 &\left[K-K_c(\sigma=0)\right]^{\frac{\tau-1}{2-\tau}} \mbox{ $\:\:\;$ if $ \tau>\frac{3}{2}$\,.}
       \end{array} \right.
              \label{eq:secondaconfrrpue}
\end{equation}
The critical behaviour of the pure model is, in general, different with respect to the case $\sigma^2>0$.
The only critical exponent that has the same value in the random field model and in 
the pure model is $\beta$, as we will show in the sequel.
We will further see that eqs. (\ref{eq:primaconfrr}) and (\ref{eq:secondaconfrr}) imply that the 
critical exponents for $\sigma^2 > 0$ do not depend on the direction that the critical line is crossed.
Moreover, the critical exponents for $T >0$ present the same values
as in the regime $T=0$. This is a well established property of random field systems, whose reasons 
can be found using renormalization group arguments \cite{bray1985scaling}.


\subsection{Calculation of $\beta$}
The critical exponent $\beta$ is defined, in the limit $h \rightarrow 0^{+}$, by the vanishing of the magnetisation 
as the critical line is approached from the ferromagnetic phase \cite{Baxter}.
Let us consider the case where $T$ is held at a fixed value
and $\sigma^2$ is the control parameter.
Using eq. (\ref{eq:jac_sigmact1}), equation (\ref{eqstate}) reduces, in the
ferromagnetic phase, to the form 
\begin{equation}
m^2=2g''(0)[\sigma_c^2(T)-\sigma^2] \,,
\end{equation}
from which we obtain that, for $T \geq 0$, the magnetisation vanishes as
\begin{equation}
m\sim \sqrt{\sigma_c^2(T)-\sigma^2}\,.
\label{eqbeta111}
\end{equation}
In the above derivation, we have used that $g''(0)$ is finite for $\tau<3/2$, as can be seen from the results shown in the appendix.
We can also study the behaviour of $m$ as a function of $T$ for fixed
$\sigma^2$. Using the fact that
\beq
\beta_c(\sigma)=\frac{2 g'(0)}{1+2\sigma^2g''(0)}\:,
\label{eq:jac_kcsigma_rand112}
\eeq
eq. (\ref{eqstate}) reduces, in the ferromagnetic phase, to
the form
\begin{equation}
m^2=\frac{1+2\sigma^2 g''(0)}{\beta}[\beta-\beta_c(\sigma)]\,,
\end{equation}
from which we have that, for $\sigma^2 \geq 0$,  the magnetisation vanishes 
according to
\begin{equation}
m\sim \sqrt{\beta-\beta_c(\sigma)}\:,
\label{eqbeta222}
\end{equation}
where we have used again that $g''(0)$ is finite for $\tau<3/2$.
From equations (\ref{eqbeta111}) and (\ref{eqbeta222}), we obtain the critical
exponent $\beta=1/2$. This is also the value 
of $\beta$ in the pure model, as can be noted from  eq. (\ref{eqbeta222}).


\subsection{Calculation of $\gamma$}

The critical exponent $\gamma$ is defined from the divergence of the 
zero-field susceptibility, defined as $\chi=\lim_{h \rightarrow 0^+} \partial m / \partial h$, as we
approach the critical line from the paramagnetic phase \cite{Baxter}.
The magnetisation is given by $m=h/z$, so that 
\begin{equation}
\chi=\frac{1}{z_0} - \lim_{h\to 0^+} \frac{h}{z^2} \frac{\partial z}{\partial h} \,,
\end{equation}
where 
$z$ here is the short hand notation for $z^{*}(h)$, namely the solution
of eq. (\ref{eqstate}) for $h > 0$.
In the paramagnetic phase, $z_0$ has a finite value that vanishes as we
approach the critical line. Using the equation of state, it can be shown that 
\begin{equation}
\frac{h}{z^2}  \frac{\partial z}{\partial h} = \frac{2Kh^2}{2Kh^2/z^3-g''(z)-z^2\beta^2g'''(z)}\:.
\end{equation}
From the results of the appendix, the above
equation vanishes for $h \rightarrow 0^{+}$ within the paramagnetic phase,
which implies that $\chi=1/z_0$. Thus, by considering the behaviour of $z_0$ close
to the critical line, we derive the following results for the critical
exponent $\gamma$
\begin{equation}
\gamma = \left\{ \begin{array}{cl}
 1 & \mbox{ if $\tau<\frac{4}{3}$} \\
  \frac{\tau-1}{3-2\tau}  & \mbox{ if $  \frac{4}{3}<\tau<\frac{3}{2} $}
       \end{array} \:.\right.
       \label{eqgammarfasa}
\end{equation}
Equation (\ref{eqgammarfasa}) holds whatever direction we choose to cross the 
critical line, including the case of $T=0$.
A similar analysis can be done for the pure model. Using eq. (\ref{eq:secondaconfrrpue}), we get the 
result obtained in reference \cite{mcguire1973spherical}:
\begin{equation}
\gamma = \left\{ \begin{array}{cl}
 1 & \mbox{ if $\tau<\frac{3}{2}$} \\
  \frac{\tau-1}{2-\tau}  & \mbox{ if $  \tau>\frac{3}{2} $}
       \end{array} \:.\right.
       \label{eqgammapureoie}
\end{equation}


\subsection{Calculation of $\alpha$}
The critical exponent $\alpha$ is defined, for $h = 0$, from the divergence of the specific heat at the critical line \cite{Baxter}. 
If $u(m,h,T)$ denotes the energy density as a function of $T$ at fixed $\sigma^2$, then $\a$ can be computed from
\begin{equation}
u(m,0, T_c(\s)+\e/2)-u(m,0,T_c(\s)-\e/2)\sim |\e|^{1-\alpha}\:,
\end{equation}
with $|\epsilon| \rightarrow 0$.
In the case where the energy density is a function of $\s$
for fixed $T$, the exponent $\a$ is defined in an
analogous way. The function $u$ is calculated above and below the critical line 
from the free energy as $u= \partial (\beta f)/ \partial \beta$.
Dropping the dependency on $T$ or $\s$, we use eqs. (\ref{freeenerffff}) and (\ref{freefinal}) to obtain
\begin{equation}
u(m,h)=u_{pure}(m,h)-\sigma^2 g' \left(\frac{h}{m} \right) -\beta \sigma^2 g''\left(\frac{h}{m}\right) \frac{\partial z_0}{\partial \beta} \,,
\label{upm}
\end{equation}
where $u_{pure}= \partial (\beta f_{pure})/ \partial \beta$. Note 
that $u_{pure}$ is not the energy density of the pure model, since the saddle-point value of $m$ depends on $\s^2$.
Equation (\ref{upm}) allows us to compute $u$ above and below
the critical line, employing the results from the appendix
combined with eqs. (\ref{eq:primaconfrr}) and (\ref{eq:secondaconfrr}). 
As a result, we obtain the values of the critical exponent $\a$: 
\begin{equation}
\alpha = \left\{ \begin{array}{cl}
 0 & \mbox{ if $\tau<\frac{4}{3}$} \\
  \frac{4-3\tau}{3-2\tau}  & \mbox{ if $  \frac{4}{3}<\tau<\frac{3}{2} $}
       \end{array} \:.\right.
       \label{eqalpharfiasia}
\end{equation}
This result holds whatever direction we cross the critical line, including
the case of $T=0$.
A similar analysis, together with  eq. (\ref{eq:secondaconfrrpue}), leads to the 
following result for the pure model
\begin{equation}
\alpha = \left\{ \begin{array}{cl}
 0 & \mbox{ if $\tau<\frac{3}{2}$} \\
  \frac{3-2\tau}{2-\tau}  & \mbox{ if $  \tau>\frac{3}{2} $}
       \end{array} \:.\right.
       \label{eqalphapurei}
\end{equation}


\subsection{Calculation of $\delta$}
The critical exponent $\delta$ is obtained 
from the vanishing of the magnetisation 
as a function of $h$ at the critical line according
to $m\sim h^{1/\delta}$ \cite{Baxter}.
From eqs. (\ref{eqstatfinalm}) and (\ref{eq:jac_eqforthecrtil}),
we obtain that $h^2 \sim z_0^3$ if $\tau <4/3$, while $h^2 \sim z_0^{1/(\t-1)}$ if $\t > 4/3$. 
Thus, eqs. (\ref{eq:primaconfrr}) and (\ref{eq:secondaconfrr}) lead to the results 
\begin{equation}
\delta = \left\{ \begin{array}{cl}
 3 & \mbox{ if $\tau<\frac{4}{3}$} \\
  \frac{1}{3-2\tau}  & \mbox{ if $  \frac{4}{3}<\tau<\frac{3}{2} $}
       \end{array} \:,\right.
       \label{eqdeltaaaa}
\end{equation}
whatever direction we choose to cross the critical line. For the pure model, a similar analysis combined with 
eq. (\ref{eq:secondaconfrrpue}) yields
\begin{equation}
\delta = \left\{ \begin{array}{cl}
 3 & \mbox{ if $\tau<\frac{3}{2}$} \\
  \frac{\t}{2-\tau}  & \mbox{ if $  \tau>\frac{3}{2} $}
       \end{array} \:.\right.
       \label{eqdeltaaaapure}
\end{equation}


\section{Considerations on the critical exponents} \label{seccomments}

From the results presented above, it is clear that $\tau \in (1, 3/2)$
is the interesting interval of $\tau$. For $\tau>3/2$, no phase transition occurs when $\sigma^2 > 0$.
The threshold
$\tau=3/2$ is called the lower critical value.
The value $\tau=4/3$, referred to as the upper critical value, plays a central role: for $\tau \in (1, 4/3)$, the mean-field
theory is valid and the critical exponents assume their classical values, whereas
for $\tau \in (4/3,3/2)$ the critical exponents are non-trivial, in the sense
that they are, in general, functions of $\tau$.
In the pure model, the upper and lower critical values are given, respectively, by $\tau=3/2$
and  $\tau=2$. We remark that the same lower and upper values of $\tau$ have been found
in the hierarchical model with Ising spins in the pure
case  \cite{kim1977critical} as well as in the presence
of random fields \cite{ParisiRoc,rodgers1988critical}.  This situation is different from 
the $D-$dimensional short-range counterpart, where the spherical model and the model with Ising spins 
have different lower critical dimensions, given by $D=4$ and $D=3$, respectively \cite{imryma,hornreich1982thermodynamic}.

Differently from the spherical model, it is commonly not possible to 
derive, within the non-mean-field region, analytical expressions for the critical exponents
in the Ising counterpart of the present model.
An exception is the exponent $\delta$. Let us define $\delta_{Pure}(\tau)$ and $\delta_{RF}(\tau)$ as the exponents
for the pure and the random field model, respectively.
In the hierarchical model with Ising spins \cite{kim1977critical, ParisiRoc}, the exponents $\delta_{RF}(\tau)$ and $\delta_{Pure}(\tau)$ have
exactly the same values as presented in eqs. (\ref{eqdeltaaaa}) and (\ref{eqdeltaaaapure}), respectively. 
Thus, one can conclude that the relation $\delta_{RF}(2-1/\tau)=\delta_{Pure}(\tau)$ is fulfilled
in the respective non-mean-field regimes of hierarchical models
with spherical as well as with Ising spins.

It is a natural question to ask if a similar mapping between the critical
properties of the pure and the random field case holds for the other critical exponents. 
In the hierarchical model with Ising spins, it has been shown, using perturbation
theory, that the relation  $\gamma_{RF}(2-1/\tau)=\gamma_{Pure}(\tau)$  
holds near the corresponding upper critical values 
of $\tau$, but fails in the non-mean-field region \cite{ParisiRoc}. 
In the spherical model, instead, this is not the case.
From the results of the previous section, we see that this mapping
is satisfied for all critical exponents, for any value of $\tau$ in the non-mean field sector.

It has been suggested that the critical properties of one-dimensional long-range 
models and $D$-dimensional short-range ones can be connected through the relation 
\cite{leuzzi2008dilute,katzgraber2009study,banos2012correspondence,leuzzi2013imry,MariaChiara}
\begin{equation}
D=\frac{2}{\tau-1},
\label{spectraldimtuaaaa}
\end{equation}
which gives the equivalent dimension $D$ of the short-range model with the same critical behaviour as the one-dimensional 
long-range model parametrised by the interaction potential $J(r)\sim r^{-\tau}$.
For models with Ising spins, such relation breaks down in the non-mean-field region \cite{ParisiRoc,MariaChiara}.
In contrast, for models with spherical spins, one can substitute $\tau$ in terms of $D$ on the critical
exponents obtained in the previous section: the resulting expressions are the same as
those derived in references \cite{mcguire1973spherical} and \cite{hornreich1982thermodynamic}, showing
that in this case there is an exact mapping between the critical behaviour 
of the hierarchical model
and that of $D$-dimensional short-range systems.
A possible reason for this can be found in reference \cite{MariaChiara}, where it 
has been shown that, under a certain super-universality hypothesis, a different relation between $D$ and $\t$ can be 
derived, which should improve eq. (\ref{spectraldimtuaaaa}) and extend its validity to the non-mean-field regime. This 
relation involves the critical exponent $\eta_{SR}(D)$, such that eq. (\ref{spectraldimtuaaaa}) is recovered for $\eta_{SR}=0$. 
Since an explicit computation in the $D$-dimensional spherical model leads to $\eta_{SR}=0$, equation (\ref{spectraldimtuaaaa})
holds exactly in this case.


\section{Conclusion} \label{secconclus}

We have studied the equilibrium properties of a spherical version of the Dyson hierarchical model in the presence of 
random fields using two independent methods: a recursive  computation of the partition
function, based on a renormalization-like transformation, and the 
standard replica approach. Both methods give exactly the same
free-energy and the same equation of state, from which it follows that the
model undergoes a paramagnetic-ferromagnetic phase transition on
the $(\sigma^2,T)$ plane, with $\sigma^2$ denoting the variance of the
random fields and $T$ the temperature. By tunning a parameter
$\t$, responsible for controlling the power-law decaying interactions, the 
hierarchical model interpolates smoothly between a mean-field and a non-mean-field
regime. We have computed analytically the critical exponents in both
regimes and their values do not 
depend on the direction that the critical line is crossed
on the phase diagram.

Two interesting results emerge from the calculation of the critical exponents.
First, there is an exact mapping between the critical behaviour of
the pure model and that of the random field model. 
In fact, we have shown that, contrary to the Ising version of the present model \cite{ParisiRoc}, the relation 
$y_{RF}(2-1/\tau)=y_{Pure}(\tau)$ holds in the whole non-mean-field sector, where $y$ denotes one of the critical
exponents considered here. Such relation, which has been proposed in 
reference \cite{ParisiRoc}, plays the role of the dimensional reduction rule
for one-dimensional long-range systems.
Second, there is an exact mapping, given by eq. (\ref{spectraldimtuaaaa}), between the critical properties of the spherical hierarchical
model in the presence of random fields and the corresponding $D$-dimensional model with
short-range interactions. This conclusion follows from the comparison of
our results for the critical exponents with those of 
references \cite{mcguire1973spherical,hornreich1982thermodynamic}. In contrast to the Ising
version of the hierarchical model, eq. (\ref{spectraldimtuaaaa}) is valid here for any value of $\tau$.

Finally, from the local stability analysis of the replica symmetric solution
we have shown that the model does not display a spin-glass phase.
Although the emergence of spin-glass states through a discontinuous
transition can not be definitely excluded, the absence of a
spin-glass phase in this model is also reinforced by the fact that the 
free-energy obtained from the recursive approach is precisely the same
as the replica symmetric free-energy. 
This result extends those of references \cite{krzakala2010elusive,krzakala2011no}, where it has 
been shown rigorously that random field systems composed of Ising spins \cite{krzakala2010elusive} or 
defined in terms of a scalar field theory \cite{krzakala2011no} do not exhibit a spin-glass phase.

The present paper opens some interesting perspectives of future works. Since 
spherical models are recovered as the $m\rightarrow \infty$ limit of $O(m)$ 
vectorial models in the pure case \cite{stanley1968spherical}, it would
be interesting to investigate perturbatively how eq. (\ref{spectraldimtuaaaa}) is modified 
in vectorial models with $m$ very large, but finite. Such study could lead to 
additional insights on the mechanism at work in the breakdown of 
eq. (\ref{spectraldimtuaaaa}), which generally occurs in low dimensional systems.
We point out that the connection between $O(m)$ vectorial models and spherical
models holds in the pure case, but it is not obvious in systems with quenched disorder. 
Thus, our results on the critical exponents of the random field model and on the absence of a 
spin-glass phase may not trivially hold for the hierarchical model with vectorial spins in the limit of a large number of components.
Another interesting perspective is the extension of our work 
to hierarchical spherical models with random couplings.
We leave this for future work.


\ack

The authors thank Giorgio Parisi and Federico Ricci-Tersenghi
for fruitful discussions and interesting suggestions.
P.U. thanks the Physics Department of the University of Rome ``La Sapienza''
where part of this work has been developed.
The research leading to these results has received funding from the European Research Council
(ERC) grant agreement No. 247328 (CriPheRaSy project).
F.L.M acknowledges the support from the Brazilian agency
CAPES through the program Science Without Borders.  
 P.U. acknowledges the support from the ERC grant NPRGGLASS.


\appendix

\section{Some useful results to calculate the critical exponents}

In this appendix we study the function $g'(z)$ and its derivatives for small 
values of $z$, which are important in the evaluation of the critical exponents.
Moreover, we compute $g'(0)$ and $g''(0)$, which are needed in the computation of 
the critical line, see eq. (\ref{eq:jac_eqforthecrtil}).

The function $g(z)$ is defined in eq. (\ref{gz}), and its
derivative $g'(z)$ reads
\begin{equation}
g'(z)=\frac{1}{2} \int d\lambda \frac{\rho(\lambda)}{z+\lambda_{\infty}-\lambda} =\frac{1}{2} 
\lim_{n\rightarrow \infty} \sum_{p=1}^{n}\frac{2^{-p}}{z+\lambda_{\infty}-\lambda_p}\:.
\label{series80}
\end{equation}
We note that $\l_\infty-\l_p$ is a positive quantity. This means that the 
function $g'(z)$ contains a sequence of simple poles on the negative part of the real $z$ axis and the 
point $z=0$ is an accumulation point for such poles. In what follows we 
assume $z\geq 0$, because we know from the equation of state that this is the physical relevant case. 
For $z>0$, the series in eq. (\ref{series80}) is always convergent because the ratio between the $p+1$-th and the $p$-th 
term is smaller than one.
For $z=0$, the series converges to the value
\beq
g'(0)=\frac{1}{2^{\t}(\l_\infty-\l_1)}\left[\frac{1}{1-2^{-(2-\t)}}-1\right] \,,
\eeq
provided  $\t<2$. 

We want to understand the behaviour of the above series for $\t > 2$, in 
the regime of small and positive $z$. Let us define $\p(z)$ as the value of $p$ that satisfies the following equation
\beq
\frac{z}{\l_\infty-\l_{\p(z)}}=1.
\eeq
Defining $\tilde p(z)=\lfloor \p(z)\rfloor$, where $\lfloor x \rfloor$ denotes 
the largest integer not greater than $x$, the behaviour of $\tilde p$, for $z\to 0$, is
such that $2^{-(\t-1)\tilde p}\sim z$.
Thus we obtain
\beq
\sum_{p=1}^\infty \frac{2^{-p}}{z+\l_{\infty}-\l_p}=\sum_{p=1}^{\tilde p}\frac{2^{-p}}{z+\l_{\infty}-\l_p}+\sum_{p=\tilde p+1}^{\infty}\frac{2^{-p}}{z+\l_{\infty}-\l_p} .
\eeq
The first term in the sum is given by
\beq
\sum_{p=1}^{\tilde p}\frac{2^{-p}}{z+\l_{\infty}-\l_p}=c_1z^{\frac{2-\t}{\t-1}} + \mathcal O(z),
\eeq
while the second one reads
\beq
\sum_{p=\tilde p+1}^{\infty}\frac{2^{-p}}{z+\l_{\infty}-\l_p}\leq c_2 z^{\frac{2-\t}{\t-1}}.
\eeq
Thus, for $\t>2$ and $z\to 0$, we get
\beq
g'(z)\sim A z^{\frac{2-\t}{\t-1}}\,,
\eeq
with $c_1$, $c_2$ and $A$ representing positive constants.

Let us now consider $g''(z)$
\beq
g''(z)=-\frac{1}{2}\sum_{p=1}^\infty \frac{2^{-p}}{(z+\l_\infty-\l_p)^2}.
\eeq
For $\t<3/2$ the series  is convergent. In particular, we have
\beq
g''(0)=-\frac{1}{2^{2\t-1} (\l_\infty-\l_1)^2}\left[\frac{1}{1-2^{-(3-2\t)}}-1\right].
\eeq
As before, we want to study the case of $\t>3/2$, with $z$ positive and small. 
The function $g''(z)$ is always convergent for $z>0$, but we expect a divergence as $z\to 0$ when $\t>3/2$.
By repeating the same argument as for $g'(z)$, in this situation we derive the expression
\beq
g''(z)\sim B z^{\frac{3-2\t}{\t-1}} \,.
\eeq
On the same lines we can get the behaviour of $g'''(z)$
\beq
g'''(z)=\sum_{p=1}^\infty \frac{2^{-p}}{(z+\l_\infty-\l_1)^3}.
\eeq
The above series is always convergent. For $\t<4/3$, we have a finite value 
for  $g'''(0)$, while we obtain that, for $z\to 0$ and $\t>4/3$, the behaviour of $g'''(z)$ is given by
\beq
g'''(z)\sim C z^{\frac{4-3\t}{\t-1}}\:.
\eeq
To summarise, we have the following asymptotic behaviours 
\begin{itemize}
\item  $ \lim_{z\rightarrow 0} g'(z) \sim z^{\frac{2-\tau}{\tau-1}}$ for $\tau > 2$, while it is finite for $\tau < 2$;
\item $  \lim_{z\rightarrow 0} g''(z) \sim z^{\frac{3-2\tau}{\tau-1}} $ for $\tau > \frac{3}{2}$, while it is finite for $\tau < \frac{3}{2}$;
\item $ \lim_{z\rightarrow 0} g'''(z) \sim z^{\frac{4-3\tau}{\tau-1}}  $ for $\tau > \frac{4}{3}$, while it is finite for $\tau < \frac{4}{3}$. 
\end{itemize}

\section*{References}
\bibliographystyle{ieeetr} 
\bibliography{bibliography}

\end{document}